# Immunization for complex network based on the effective degree of vertex

Ke Hu and Yi Tang[*]

*Department of Physics and Institute of Modern Physics, Xiangtan University, Xiangtan 411105, Hunan, China*

The basic idea of many effective immunization strategies is first to rank the importance of vertices according to the degrees of vertices and then remove the vertices from highest importance to lowest until the network becomes disconnected. Here we define the effective degrees of vertex, i.e., the number of its connections linking to un-immunized nodes in current network during the immunization procedure, to rank the importance of vertex, and modify these strategies by using the effective degrees of vertices. Simulations on both the scale-free network models with various degree correlations and two real networks have revealed that the immunization strategies based on the effective degrees are often more effective than those based on the degrees in the initial network.

## 1. Introduction

The epidemic dynamic on complex network is largely benefited from the need of understanding the spreading of both computer viruses and human sexual diseases[1-3], which has recently become a subject of intensive investigations.[4-15] Since most of real networks involving the Internet and the web of sexual contacts appear to be of large heterogeneity in the connectivity properties,[16-19] epidemic modeling deals with the complex systems with heterogeneous connectivity to approach a more realistic situation. This highly heterogeneous topology of the networks is mainly reflected in the small average path lengths among any two nodes (small-world property)[20], and in a power law connectivity distribution (scale-free property)[19]. These properties are known to enhance the spread of diseases, making them harder to eradicate, which calls for specific immunization strategies.[4-11]

---

*Corresponding author. Email address: tangyii@yahoo.cn

During the past few years, a number of immunization strategies have been proposed, aiming to lower the immunization fraction of the networks for economical application.[21-24] Among them, random immunization has been shown to be disabled when connection distribution of the networks is wide and scale-free, since it needs to immunize almost all of vertices to eradicate the epidemic[21]. A most efficient strategy for immunization is the targeted immunization (TI) which is defined based on the vertices' hierarchy. In this strategy, an epidemic can be prevented by immunizing just a small amount of the most highly connected vertices.[21] However, good knowledge of degrees of each vertex in the networks is required. Such global knowledge about networks often proved to be difficult to obtain, which results in that this approach is unrealistic for many practical applications. From this point of view, Cohen *et al* [22,23] devised a strategy, named as acquaintance immunization, which integrates both efficiency and advantage of being purely local. According to this strategy, random acquaintances of random vertices are required to be immunized. Therefore no knowledge of the vertices' degrees and any other global knowledge are required. Its efficiency greatly exceeds that of random immunization while less efficient than targeted immunization. Another more effective immunization strategy, called as enhanced acquaintance immunization (EAI), is recently introduced[24]. It is a local strategy. Moreover, its efficiency is proved to approach to that of the targeted one. More recently, a novel immunization based on the equal graph partitioning is proposed[25]. Although it requires a global knowledge about network, it is significantly better than the TI strategy. Owing to the immediate practical and economical implications, further studies involving the improvement of immunization and the other effective strategies is still of considerable significance.

In general, immunization of a vertex is regarded as removal of both this vertex and its connections. The goal of the immunization process is to pass (or at least approach) the immunization threshold, leading to minimization of the number of infected vertices. Because the connections of an immunized vertex are synchronously removed, immunization of each vertex will directly leads to the reduction of degrees of its neighbors. Therefore, during the immunization procedure, immunization according to the hierarchy of degrees of vertices in the initial network is not always most efficient in some cases. Here, we define the effective degree of a vertex as the number of its connections linking to un-immunized vertices. From the networks displayed in Fig. 1, it is clear that immunizing vertex 8 is more efficient than that to immunizing vertex 4 despite that the degrees of vertex 8 is smaller than that of vertex 4. Thus we suggest that the previous strategies should be modified by recalculating the degrees of vertices in the current network during the immunization procedure. In addition, it should be pointed out that this modification can retain the advantage of being purely local of the previous local strategies since estimating the effective degrees of a vertex requires only a local knowledge: the number of its neighbors that are not immunized.

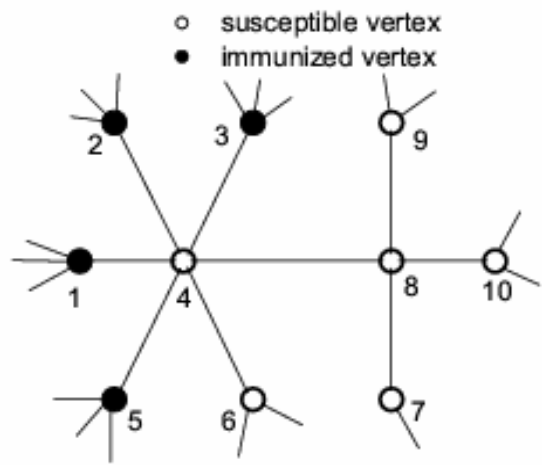

Fig. 1. Example network used for illustrating the immunization scheme based on the effective degrees.

## 2. Modification for the strategies

Based on the definition of the effective degrees, one can in principle apply the present scheme to modify any degree-based immunization strategies. Here we will focus on two typical degree-based immunization strategies. One is the TI which is believed to be most efficient among all the degree-based strategies although it is a global one and requires a complete knowledge of the degrees of the vertices. We modify the TI strategy by progressively making immune the vertices with highest effective degrees. Another is the EAI, where two variations was studied[24]: 1. the selected node points randomly to a neighbor which is more connected than himself, then the neighbor is immunized. If there are no such neighbors, no vertex is immunized; 2. selecting a random vertex and asking for a neighbor that has more degrees than a given threshold $k_{cut}$ and immunizing it. For simplicity we denote the two variations by EAI-1 and EAI-2 respectively in the rest of this paper. The two immunization conditions corresponding these two variations may be modified by using the effective degrees: 1. if the effective degree of a neighbor of the selected vertex are larger than the degree of the selected vertex, the neighbor is immunized, otherwise, no vertex is immunized; 2. selecting a random vertex and asking for a neighbor which its effective degrees are more than a given threshold $k_{cut}$ and immunizing it.

## 3. Tests of the modified strategies

In the presence of the immunization, we study the cost effectiveness by looking at the infection's prevalence in the stationary regime (endemic state) as a function of the fraction of immunized vertices $f$. Initially we infect half of the susceptible vertices, and iterate rules of the SIS model with parallel updating. After a transient regime the system reaches a steady state, characterized by a constant average density of infected nodes, i.e., the infection's prevalence $\rho_f$ which is the time average of the fraction of infected nodes. The reduced prevalence is defined as the rate $\rho_f/\rho_0$, where $\rho_0$ is the infection's prevalence without immunization.[26] A critical fraction $f_c$ of immunization above which the prevalence in the steady state is null, is referred to the immunization threshold.

Many realistic complex systems possess, along with the SF property, some more detailed topological features which vary strongly for different systems.[27-29] Due to the lack of a single model encompassing the topological features of complex networks, we consider a few SF networked models and two real networks, aiming to test the modified immunization strategies.

### 3.1 The Barabási and Albert scale-free network

As a paradigmatic example of SF networks, we consider the Barabási and Albert (BA) model which has been universally used as a general structure to study various dynamical processes taking place in complex networks.[30] This model introduces two simple mechanisms that are believed to be common in reality: growth and preferential attachment, to construct the SF networks.[19] We follow the algorithm devised in Ref.[19]: Starting with a small number $m_0$ of nodes; and then at every time step a new node is added, with $m$ links that are connected to an old node $i$ with $k_i$ links according to the probability $k_i/\sum_j k_j$. After iterating this scheme a sufficient large number of times, we obtain a network composed of $N$ nodes with connectivity distribution $P(k) \propto k^{-3}$ and average connectivity $<k>=2m$.

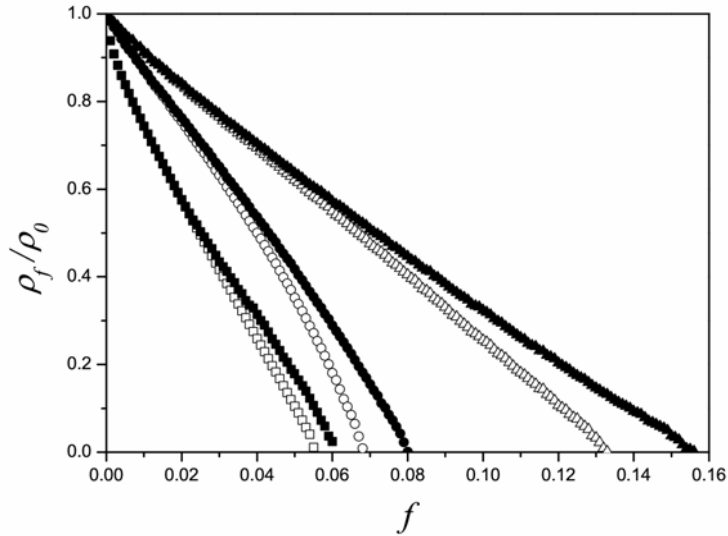

Fig. 2. Reduced prevalence $\rho_f/\rho_0$ from computer simulations of the SIS model in the BA networks with the TI (square), the EAI-1(up triangles) and the EAI-2 (circles) with a given threshold $k_{cut}=11$, at a fixed spreading rate $\lambda=0.25$. The full symbols correspond to the original strategies, while the empty symbols correspond to the modified strategies. The prevalence is computed averaging over at least 100 different starting configurations, performed on at least 10 different realizations of the networks. The network size $N=10^5$ and the average degree $<k>=6$.

Numerical simulations for the epidemic spreading on the BA model have been performed to estimate the efficiencies of the present scheme. Figure 2 shows that the reduced prevalence $\rho_f/\rho_0$ for both modified and original strategies suffers a very sharp drop and exhibits the onset of an immunization threshold above which no endemic state is possible. However, for the strategies modified by the present scheme, the immunization thresholds are lower than that for the corresponding original ones. These results indicate that on the BA networks, the present scheme could further improve these strategies.

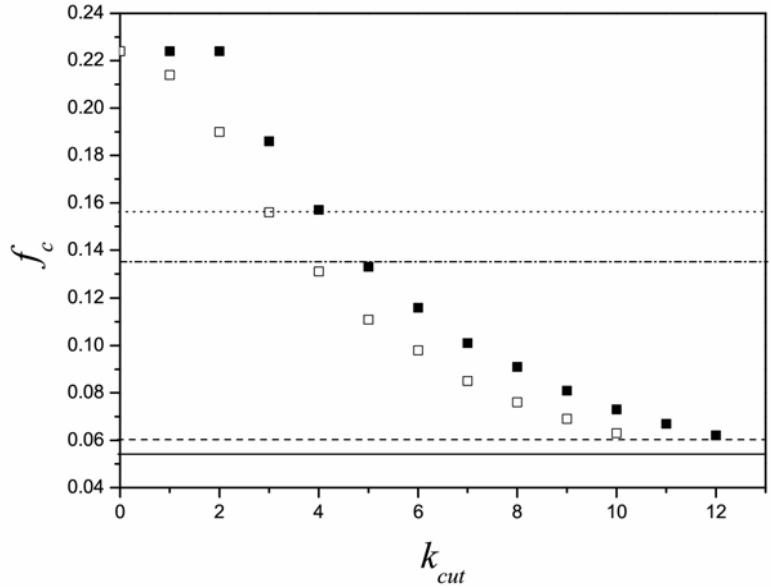

Fig. 3. Critical immunized fraction $f_c$ as a function of the threshold value $k_{cut}$ for the EAI-2 strategy applied to the BA networks with the average degree $<k>=6$, and systems size $N=10^5$. The full symbols correspond to the EAI-2, while the empty symbols correspond to the modified EAI-2. The upper horizontal dotted line and dash-dotted line is the result for the EAI-1 and the modified EAI-1, the dashed line in the middle corresponds to the TI, while the lower line refers to the modified TI.

It should be pointed out that in the second immunization condition of the EAI, the given threshold $k_{cut}$ should be estimated before immunization. Usually, it is very difficult to obtain an optimal $k_{cut}$ since it depends on the system's size, the scaling exponent of power-law degree distribution and the spreading rate, even some other factors. If $k_{cut}$ is too large, immunizing all vertices with degree $k>k_{cut}$ is still not prevent the prevalence of infection. If too small, an optimal immunization threshold can not be reached. Due to this limit, by applying the present scheme to the EAI, the lowest immunization threshold is not still lower than that of the TI. However, the present scheme can relax the estimation for $k_{cut}$, allows a lower $k_{cut}$ to obtain low immunization threshold. In figure 3, we give the immunization threshold as a function of $k_{cut}$. When $k_{cut}$ is increased, the immunization threshold decreases and finally approaches to the optimal critical fraction of immunization $f_c(\approx0.06)$. For each value of $k_{cut}>0$, the immunization threshold for the modified EAI is lower than that for the original EAI.

In addition, as is worth mentioning, the ability that the present scheme improves the previous immunization strategies is dependent on the order of immunization of vertices. Perhaps in a proper order, a greater improvement for the previous strategies can be reached. However, immunizing vertices follow such a certain proper order requires a complete acknowledge of networked structures which often proves to be difficult to gather. Thus in our simulations, the order for immunization is not considered.

### 3.2 Assortative and disassortative networks

Many realistic networks have been found to exhibit assortative or disassortative mixing patterns of degree which also are referred to the degree correlations.[27] For instance, many social networks exhibit that the nodes with high connectivity (hubs) connect more preferably to the hubs, which are called as the assortatively mixed networks. On the opposite side, many technological and biological networks show disassortative mixing, i.e., the hubs are preferably connected to the nodes with small degree. In this stage, the BA model that has been used as a model of the structures of many realistic networks is incomplete due to the absence of these mixing patterns.[27] Therefore, testing the present immunization strategies on the assortatively and disassortatively mixed SF topologies would be a better approach to the realistic situation than that on the BA model.

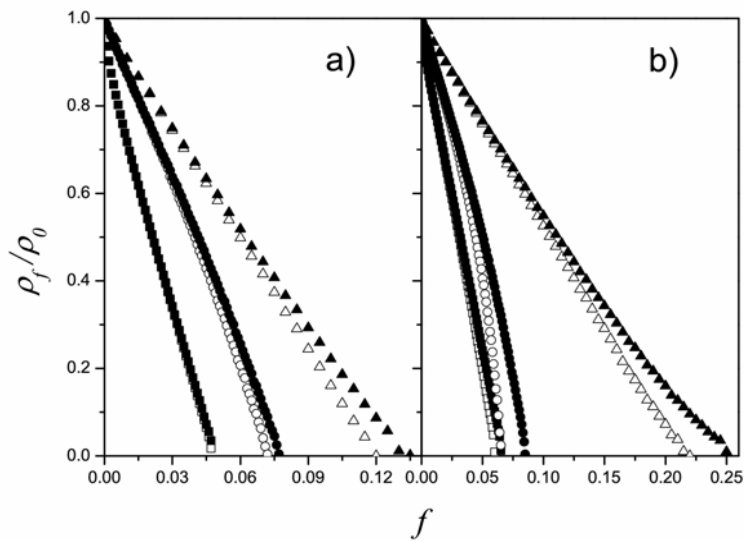

Fig.4. Reduced prevalence $\rho_f/\rho_0$ from computer simulations of the SIS model in the disassortatively (a) and assortatively (b) mixed networks with the TI (squares), the EAI-1 (up triangles) and the EAI-2 (circles) with a given threshold $k_{cut}=9$, at a fixed spreading rate $\lambda=0.25$. The full symbols correspond to the original strategies, while the empty symbols correspond to the modified strategies. The network's parameters are the same as in Fig. 1.

Complex networks with the assortative or disassortative mixing can be constructed by applying a Metropolis dynamics to a random graph with the desired degree distribution.[27] Especially, we here regard the BA model as an initial graph to ensure that all networks possess same power-law degree distribution. For detailed algorithm for generating these networks please see Ref. [27]. In each realization of the networks, we obtain the assortatively and disassortatively mixed networks after $5\times10^5$ Monte Carlo steps, and the Pearson correlation coefficients[27] averaging over 10 different realizations of the networks are equal to 0.0579 and -0.0441, respectively. Previous studies have shown that the assortatively mixed networks are more robust to removal of the hubs,[27] so that the degree-based immunization strategies seem to be somewhat ineffective, while the disassortatively mixed networks are more vulnerable to removal of their hubs,[27] and thus, these strategies should be more effective for such networks. However, on the assortative network, the present scheme is more efficient to improve the previous strategies than that on the disassortative network, i.e., greater improvement is obtained on the assortative network while is null on the disassortative. This difference of the efficiency is rooted in the mixing patterns. Intuitively, the immunization of vertices can reduce both the average degree of remain sub-network and the maximum connectivity $k_{max}$ which have the effect of restoring a bound in the connectivity fluctuations, inducing an effective epidemic threshold. When progressively immunize the hubs, the average degree of the remain network would be reduced more on the disassortatively mixed networks than that on the assortatively mixed networks, while the maximum connectivity would be lower on the assortatively mixed networks than that on the disassortatively mixed ones. From this point of view, it can be expected that on the assortatively mixed networks, the present scheme can obtain a greater improvement of the efficiencies of these immunization strategies than that on the disassortatively mixed networks despite the assortative networks are more prone to the spreading and persistence of infections.

## 3.3 Real-world networks

Finally, we also test the modified TI and EAI on two real networks with different mixing pattern from different fields. Figure 5 shows simulation results for the autonomous system (AS) Internet and the collaboration network of scientists in condensed matter. The real map of the AS Internet considered here, was reconstructed from BGP tables posted by the University of Oregon Route Views Project, and available at the web site http://www-personal.umich.edu/~mejn/netdata/ and contains 22963 vertices and 48436 links, corresponding to an average connectivity $<k>\approx 4.2186$. This network is very important in the study of computer virus spreading. The collaboration network is the updated network of coauthorships between scientists posting preprints on the Condensed Matter E-Print Archive and includes all preprints posted between Jan 1, 1995 and March 31, 2005.[31] Simulation is performed regardlss of the weights of coauthorships. This network has 40421 vertices and 175693 links, which leads to a high average degree of 8.6932. For the AS Internet, the advantage of the present scheme is about *1%* better compared to the original TI, *16%* for the EAI-2 with threshold $k_{cut}$=*7* and *14%* for the EAI-1. On the collaboration network, the present scheme shows an advantage of about *13%* for the TI, *29%* for the EAI-2 with threshold $k_{cut}$=*9* and *15%* for the EAI-1. This confirms that the present scheme is indeed effective to improve the previous immunization strategies on both the Internet and the collaboration network. It should be noted that the performance efficiency of the present scheme is better in the collaboration network than that in the Internet. This is due to the different mixing pattern between the two real networks: the AS Internet shows the high disassortative mixing with the Pearson correlation coefficient -0.1984, while the collaboration network shows the assortative mixing with the Pearson correlation coefficient 0.1863.

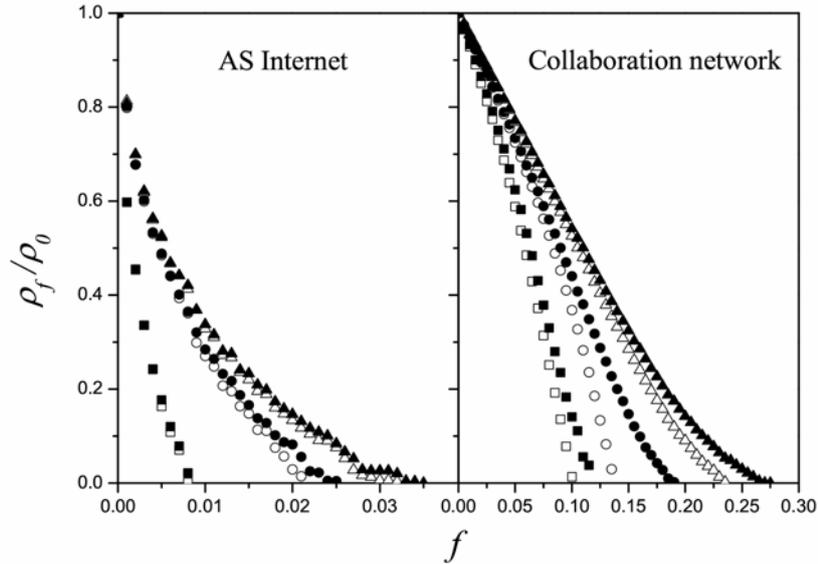

Fig.5. Reduced prevalence $\rho_f/\rho_0$ from computer simulations of the SIS model in a portion of both the Internet and the collaboration network with the TI (squares), the EAI-1 (up triangles) and the EAI-2 (circles) with a given threshold $k_{cut}=7$, at a fixed spreading rate $\lambda=0.25$. The full symbols correspond to the original strategies, while the empty symbols correspond to the modified strategies.

## 4. Summary and discussion

We have proposed a scheme to improve the degree-based immunization strategies. By comparing the efficiencies of several immunization strategies on several different SF networks, it is found that the present scheme can improve the efficiency of the previous effective immunization strategies. Especially, the improvement is unexpectedly great for the assortatively mixed networks, despite such networks are more prone to the spreading and persistence of infections.

As a final remark, the present scheme considers the degrees of vertices as a measure of the importance of vertices. Other quantities, such as the betweenness centrality[32], or integrating it with the degree can be considered to represent the importance of vertices. In addition, almost all immunization strategies are based on the immunization of vertices. Thus, in despite of the great advances in the efficiency of these strategies, they are not suitable for some epidemics where a low-cost and effective bacterin is not successfully developed. In this case, we suggest that some costless strategies can be developed based on a selective removal of edges according to the importance of edges to prevent the epidemic prevalence, or even modify the method by recalculating the importance of edges after every step of edges removal.

## Acknowledgement

This work has been supported by the General Project of Hunan Provincial Educational Department of China under Grant No. 07C754, and the National Natural Science Foundation of China under Grant No. 30570432.